\documentclass{bmcart}

\usepackage{amsthm,amsmath}
\usepackage[utf8]{inputenc} 

\usepackage{amsfonts}
\usepackage{graphicx}
\usepackage{epsfig}
\usepackage{subfigure}
\usepackage{bbm}
\usepackage{psfrag}
\usepackage{pstool}
\usepackage{amsbsy}
\newcommand{\ba}{\begin{align}}
\newcommand{\ea}{\end{align}}
\newcommand{\be}{\begin{equation}}
\newcommand{\ee}{\end{equation}}
\newcommand{\bea}{\begin{eqnarray}}
\newcommand{\eea}{\end{eqnarray}}

\def\includegraphics{}

\startlocaldefs
\endlocaldefs

\begin{document}

\begin{frontmatter}

\begin{fmbox}
\dochead{Research}


\title{Analog quantum simulation of gravitational waves in a Bose-Einstein condensate}


\author[
   addressref={aff1},                   
   corref={aff1},                       
   email={pmxtb2@nottingham.ac.uk}   
]{\inits{T}\fnm{Tupac} \snm{Bravo}}
\author[
   addressref={aff1},
   email={carlos.sabin@nottingham.ac.uk}
]{\inits{C}\fnm{Carlos} \snm{Sab{\'\i}n}}
\author[
   addressref={aff1},
   email={ivette.fuentes@nottingham.ac.uk}
]{\inits{I}\fnm{Ivette} \snm{Fuentes}}


\address[id=aff1]{
  \orgname{School of Mathematical Sciences, University of Nottingham}, 
  \street{University of Nottingham},                     %
  \postcode{NG7 2RD}                                
  \city{Nottingham},                              
  \cny{UK}                                    
}


\begin{artnotes}
\end{artnotes}

\end{fmbox}


\begin{abstractbox}

\begin{abstract} We show how to vary the physical properties of a Bose-Einstein condensate (BEC) in order to mimic an effective gravitational-wave spacetime. In particular, we focus in the simulation of the recently discovered creation of particles by a \textit{real} spacetime distortion in box-type traps. We show that, by modulating the speed of sound in the BEC, the phonons experience the effects of a simulated spacetime ripple with experimentally amenable parameters. These results will inform the experimental programme of gravitational wave astronomy with cold atoms.
%
\end{abstract}




\end{abstractbox}
%

\end{frontmatter}



\section*{Introduction}
Quantum simulators \cite{review} were originally conceived by Feynman as experimentally amenable quantum systems whose dynamics would mimic the behaviour of more inaccessible systems appearing in Nature. Together with the exploration of this visionary insight in countless quantum platforms at an increasingly accelerated rate, last years had witnessed the birth of alternate approaches to quantum simulation. For instance, a quantum simulator can be used to emulate a phenomenon predicted by a well stablished theory but very hard to test in the laboratory, such as Zitterbewegung \cite{naturekike}. Going a step further, it can also be used to materialise an artificial dynamics that has never been observed in Nature while being theoretically conceivable \cite{majorana1,majorana2} or even to emulate the action of a mathematical transformation \cite{noncausal}.

Einstein's theory of general relativity \cite{rindlerrelativity} predicts the existence of gravitational waves \cite{reviewgravwaves}, namely perturbations of the spacetime generated by accelerated mass distributions. Since sources are typically very far from Earth, the theory predicts that the amplitude of gravitational waves reaching our planet is extremely small and thus, finding experimental evidence of their existence is a difficult task. Indeed the quest for the detection of these spacetime distortions \cite{gravwavesdetectors} has been one of the biggest enterprises of modern science and the focus of a great amount of work during the last decades, both in theory and experiment. Recently, some of the authors of this manuscript have proposed a novel method of gravitational wave detection based on the generation of particles in a Bose-Einstein condensate produced by the propagation of the gravitational ripple \cite{ourgravwave}. Detection is carried out through a resonance effect which is possible because the range of frequencies of typical gravitational waves is similar to the one of the Bogoliubov modes of a BEC confined in a box-like potential. Then the gravitational wave is able to hit a particle creation resonance, in a phenomenon resembling the Dynamical Casimir Effect \cite{moore, casimirwilson}. Due to the frequencies involved, this effect is completely absent in optical cavities.  The use of  our technique would relax some of the most daunting demands of other programmes for gravitational wave detection such as the use of highly-massive mirrors and $km$-long interferometer arms. However, due to the extremely small amplitude of the gravitational waves when they reach the Earth, their detection is always challenging, since it requires an experimental setup extremely well isolated from possible sources of noise. It would be of great benefit for the experiment if the amplitude of the spacetime ripples were larger. 

In this paper, we show how to realise a quantum simulation of the generation of particles by gravitational waves in a BEC.  We exploit the fact that the Bogoliubov modes of a trapped BEC satisfy a Klein-Gordon equation on a curved background metric. The metric has two terms \cite{matt, liberati,salelites,ourgravwave}, one corresponding to the real spacetime metric and a second term, corresponding to what we call the analogue gravity metric, which depends on BEC parameters such as velocity flows and energy density. While in \cite{ourgravwave} we analyse the effect of changes in the real spacetime metric, in this case we consider the manipulation of the analogue gravity \cite{analoguereview2011, luisblackhole, megamind,jeffblackhole} metric, assuming that the real spacetime is flat. Since in this case the experimentalist is able to manipulate artificially the parameters of the condensate, we are able to simulate spacetime distortions with a much larger amplitude, as if the laboratory were closer to the source of the gravitational ripples. We show that with realistic experimental parameters, a physically meaningful model of gravitational wave can be simulated with current technology.

The paper is organised as follows. First we review the effects of a \textit{real} gravitational wave in the Bogoliubov modes of the BEC. Then we consider the case in which there are non-zero initial velocity flows in the BEC, showing that there is always a reference frame in which we can modulate the speed of sound in such a way that the phonons experience an effect analog to the one produced by the propagation of a real spacetime wave. Finally, for the sake of simplicity we assume that there are no velocity flows in the BEC and we show that the gravitational wave can be simulated with current technology.

\section*{Gravitational waves in a BEC}
The metric of a gravitational wave spacetime is commonly modelled by a small perturbation $h_{\mu\nu}$ to the flat Minkowski metric $\eta_{\mu\nu}$, i.e. \cite{reviewgravwaves}, \begin{equation}\label{eq:gmunu}
g_{\mu\nu}=\eta_{\mu\nu}+h_{\mu\nu}
\end{equation}
where 
\begin{equation}\label{eq:metricplane}
\eta_{\mu\nu}= \left(\begin{array}{cccc}-c^2&0&0&0\\0&1&0&0\\0&0&1&0\\0&0&0&1\end{array}\right).
\end{equation}
and $c$ is the speed of light in the vacuum. We consider Minkowski coordinates $(t,x,y,z)$. In the transverse traceless (TT) gauge \cite{reviewgravwaves}, the perturbation corresponding to a gravitational wave moving in the z-direction can be written as,
\begin{equation}
h_{\mu\nu}= \left(\begin{array}{cccc}0&0&0&0\\0&h_{+}(t)&h_{\times}(t)&0\\0&h_{\times}(t)&-h_{+}(t)&0\\0&0&0&0\end{array}\right),
\end{equation}
where $h_{+}(t)$, $h_{\times}(t)$ correspond to time-dependent perturbations in two different polarisations. Later on we will restrict the analysis to 1-dimensional fields, where the line element takes a simple form,
\begin{equation}\label{eq:lineelement}
ds^2=-c^2\,dt^2+(1+h_{+}(t))\,dx^2.
\end{equation}

We are interested in simulating this particular spacetime in a BEC. To this end we use the description of a BEC on a general spacetime metric following references \cite{matt, liberati}.  This description stems from the theory of fluids in a general relativistic background \cite{matt} and thus is valid as long as the BEC can be described as a fluid \cite{liberati}. In the superfluid regime, a BEC  is described by a mean field classical background $\Psi$ plus quantum fluctuations $\hat\Pi$. These fluctuations, for length scales larger than the so-called healing length, behave like a phononic quantum field on a curved metric.  Indeed, 
in a homogenous condensate, the massless modes of the field obey a Klein-Gordon equation 
\begin{equation}\label{eq:box}
\Box\hat\Pi=0 
\end{equation}
where the d' Alembertian operator 
\begin{equation}\label{eq:box2}
\Box=1/\sqrt{-\mathfrak{g}}\,\partial_{a}(\sqrt{-\mathfrak{g}}\mathfrak{g}^{ab}\partial_{b}) 
\end{equation}
depends on an effective spacetime metric $\mathfrak{g}_{ab}$ -with determinant $\mathfrak{g}$- given by  \cite{matt,liberati,salelites}
\begin{equation}\label{eq:effectivemetric}
\mathfrak{g}_{ab}=\frac{\rho\,c}{c_s}\left[g_{ab}+\left(1-\frac{c_s^2}{c^2}\right)V_aV_b\right].
\end{equation}
The effective metric is a function of the real spacetime metric $g_{ab}$ (that in general may be curved) and   
background mean field properties of the BEC such as the number density $n_0$, the energy density $ \rho_0$, the pressure $p_0$ and the speed of sound $c_s=c\sqrt{\partial p/\partial\rho}$.  Here $p$ is the total pressure, $\rho$ the total density and $V_a$ is the 4-velocity flow on the BEC. In the absence of background flows $V_a=(c,0,0,0)$ and then,
\begin{equation}\label{eq:effmetric}
\mathfrak{g}_{ab}= \frac{\rho\,c}{c_s} \left[g_{ab}+ \left(\begin{array}{cccc}(c^2-c_s^2)&0&0&0\\0&0&0&0\\0&0&0&0\\0&0&0&0\end{array}\right)\right]. 
\end{equation}
In the absence of a gravitational wave, the real spacetime metric is $g_{ab}=\eta_{ab}$, where $\eta_{ab}$ has been defined in Eq. (\ref{eq:metricplane}). Therefore, the effective metric of the BEC phononic excitations on the flat spacetime metric is given by,
\begin{equation} \label{eq:phonons}
\mathfrak{g}_{ab}= \frac{\rho\,c}{c_s}\begin{pmatrix}-c_s^2&0&0&0\\0&1&0&0\\0&0&1&0\\0&0&0&1\end{pmatrix}.
\end{equation}
Ignoring the conformal factor -which can always be done in 1D or in the case in which is time-independent- we notice that the metric is the flat Minkowski metric with the speed of light being replaced by the speed of sound $c_s$. By considering a rescaled time coordinate $t'=(c/c_s)t$ we recover the standard Minkowski metric $ds^2=-c^2dt^{2}+dx^2$. This means that the phonons live on a spacetime in which, due to the BEC ground state properties, time flows in a different fashion and excitations propagate accordingly. 

The real spacetime metric of a gravitational wave is given by Eq. (\ref{eq:gmunu}) and thus, the effective metric for the phonons is,
\begin{equation} \label{eq:phononsgw}
\mathfrak{g}_{ab}= \frac{\rho\,c_s}{c} \left(\begin{array}{cccc}-c_s^2&0&0&0\\0&1+h_{+}(t)&h_{\times}(t)&0\\0&h_{\times}(t)&1-h_{+}(t)&0\\0&0&0&1\end{array}\right).
\end{equation}
For simplicity, we considered a quasi one-dimensional BEC.  
The line element is conformal to,
\begin{equation}\label{eq:lineelement}
ds^2=-c_s^2\,dt^2+(1+h_{+}(t))\,dx^2.
\end{equation}
In \cite{ourgravwave}, it is shown that the propagation of the gravitational wave generates particles in  a BEC confined in a box-like trap. This particle creation is characterised by the Bogoliubov coefficients $\beta_{mn}=-(\hat{\phi}_m,\phi_n^*)$, where $\hat{\phi}_m$, $\phi_n$ are the $m$ and $n$ mode solutions of  Eq. (\ref{eq:box}) given by the metrics in Eq. (\ref{eq:phononsgw}) and Eq. (\ref{eq:phonons}) respectively. In particular, if we model the gravitational wave by a sinusoidal oscillation of frequency $\Omega$ that matches the sum of the frequencies of a certain pair of modes $m$ and $n$, the number of particles grows linearly in time. We refer the reader to \cite{ourgravwave} for more details.

\section*{Quantum simulation}

The aim of this section is to show how to get a line element similar to the one in Eq. (\ref{eq:lineelement}) by manipulating the parameters of the BEC while the real spacetime is assumed to be flat.

Going back to the effective metric Eq. (\ref{eq:effectivemetric}) and restricting ourselves to 1D we perform the following coordinate transformation \cite{liberati}:

\begin{equation}
\chi = \sqrt{c^2t^2-x^2} \ \ , \ \  \zeta = \frac{x}{\sqrt{c^2t^2-x^2}} \ .
\end{equation}

Next, we choose the velocity profile such that in this new coordinate system the BEC is at rest, i.e., 
\begin{equation}\label{eq:vecprofchi}
v^\chi = c\,;\, v^\zeta = 0. 
\end{equation}
The velocity profile in the original coordinate system needed in order to do this is therefore 
\begin{equation}\label{eq:vecproft}
v^t = c \sqrt{1+\zeta^2}\,;\, v^x=c \zeta.
\end{equation}
The line element given by the metric in Eq. (\ref{eq:effectivemetric}) then takes the form
\begin{equation}\label{eq: bec_metric_2}
ds^2 = -\rho(\chi) \frac{c_s(\chi)}{c}d\chi^2 + \rho(\chi) \frac{c}{c_s(\chi)}\chi^2 \frac{d\zeta^2}{1+\zeta^2} \,
\end{equation}
in which we are assuming that the speed of sound $c_s$ and the density $\rho$ might depend on the coordinates. We make another coordinate transformation (this time only in $\chi$), so that
\begin{equation}
c^2_{s0}d\tau^2 = \rho(\chi) \frac{c_s(\chi)}{c}d\chi^2  \ .
\end{equation}
Here, we denote by $c_{s0}$ the constant value of the speed of sound in the absence of any manipulation of the BEC parameters. Since we still have the freedom to specify the velocity of sound or the density of the BEC, we choose to let the density be constant, so that $\rho = \rho_0$ and we fix the velocity of sound such that
\begin{equation}\label{eq: sound_velocity}
\ell^2 (1+h_+(\tau)) \equiv \chi^2 \rho_0\frac{c}{c_s(\chi)} \ ,
\end{equation}
where $\ell$ is a constant with units of distance. In order to do this, $\chi$ as a function of $\tau$ must be
\begin{equation}\label{eq: chi_tau}
\chi(\tau) = \sqrt{\frac{2\,c_{s0}}{\rho_0}\ell \int^\tau_0\sqrt{1+h_+(s)}ds} \ .
\end{equation}
Modelling the oscillation of the $+$- polarisation of the gravitational wave in the standard way as 
\begin{equation}\label{eq:wave+}
h_+(\tau) = A_+\sin(\Omega \tau), 
\end{equation}
expanding the root in the integral in powers of $h_+$ and keeping only linear terms in $A_+$,  Equation \ref{eq: chi_tau} takes the more friendly form:
\begin{eqnarray}\label{eq:chi_tau_2}
\chi(\tau) & \approx \sqrt{\frac{2 \ell c_{s0}}{\rho_0}\int^\tau_0\left(1+\frac{A_+}{2}\sin(\Omega s)\right)ds} \nonumber \\
& =  \sqrt{\frac{2 \ell\, c_{s0}}{\rho_0}\left[\tau+\frac{A_+}{2\Omega}(1-\cos(\Omega \tau))\right]} \ ,
\end{eqnarray}






up to first order in $A_+$. With this, we can solve Eq. (\ref{eq: sound_velocity}) for the velocity of sound in terms of $\tau$ -substituting Eq. (\ref{eq:chi_tau_2}) in Eq. (\ref{eq: sound_velocity}) as 

\begin{eqnarray}
c_s (\tau) & = &\frac{c\,c_{s0}}{\ell} \left[ \frac{2\Omega \tau + A_+(1-\cos(\Omega \tau))}{\Omega(1+ A_+ \sin(\Omega \tau))}\right] \nonumber \\
& \approx& \frac{c\,c_{s0}}{\ell \Omega}(2\Omega \tau + A_+[1-\cos(\Omega \tau) - 2\Omega \tau \sin(\Omega \tau)]) \ ,
\end{eqnarray}
up to first order in $A_+$. 
Now that we obtained an explicit expression for the velocity of sound as a function of the coordinate $\tau$, it is natural to ask the significance of the constant parameter $\ell$. For this, we see that in the absence of a simulated gravitational wave, Equation \eqref{eq: sound_velocity} becomes $\ell = \chi_0 \sqrt{\rho_0c/c_{s0}(\chi_0)}$, for a particular value $\chi=\chi_0$. A convenient choice is $\ell = 2c\tau_0$, so that the velocity of sound can be expresed as $c_s(\tau_0) = c_{s0}$ for $\tau=\tau_0$. Hence, $\ell$ is represented by the hyperbola $c^2t^2 - x^2 = \ell^2c_{s0}/(\rho_0c)$ in the Minkowski coordinate system.
A final coordinate transformation is made, defining $\xi$ as 
\begin{equation}\label{eq:finalchange}
\zeta = \sinh(\frac{\xi}{\ell}),
\end{equation}
such that $d\xi^2 = \ell^2 d\zeta^2/(1+\zeta^2),$ thus transforming the line element Eq. (\ref{eq: bec_metric_2}) into 
\begin{equation}
ds^2 = -c^2_{s0}d\tau^2 + (1+h_+(\tau))d\xi^2 \ ,
\end{equation} 
as desired.


\subsection*{The case with no background flows. Experimental implementation}
In the last section, we have shown how the speed of sound can be modified to mimic a gravitational wave in a coordinate system $(\tau, \xi)$ in which there are no background flows. Now, we will consider for simplicity that this coordinate system is the lab frame $(t,x)$ and relate our results directly with experimental parameters.

As we have already seen above Eq. (\ref{eq:phonons}) -and assuming again a 1D spacetime- if $v^t=c$ and $v^x=0$, then the line element is conformal to:
\begin{equation}\label{eq:phonons2}
ds^2=-c^2_s\,dt^2+dx^2
\end{equation} 
Now, if we consider that the speed of sound can depend on $t$:
\begin{equation}\label{eq:speedofsound}
c_s(t)=c_{s0}\,f(t),
\end{equation}
the corresponding line element is conformal to:
\begin{equation}\label{eq:phonons3}
ds^2=-c^2_{s0}\,dt^2+\frac{1}{(f(t))^2}\,dx^2.
\end{equation}
Therefore, if the speed of sound varies in time such that
\begin{equation}\label{eq:cst}
f(t)=(1-\frac{A_+\,\sin{\Omega\,t}}{2}),
\end{equation} 
we find, up to the first order in $A_+$:
\begin{equation}\label{eq:cstgw}
ds^2=-c^2_{s0}\,dt^2+(1+A_+\,\sin{\Omega\,t})\,dx^2.
\end{equation}
 So the experimental task is to modulate the speed of sound as:
 \begin{equation}\label{eq:gravwavesin}
 c_s(t)=c_{s0}(1-\frac{A_+\,\sin{\Omega\,t}}{2}).  
 \end{equation}
 In a weakly interacting condensate \cite{pethicksmith}, the speed of sound is
 \begin{equation}\label{eq:speedofsound}
 c_{s0}=\sqrt{\frac{\rho\,g}{m}}
 \end{equation}
 where $\rho$ is the density, $m$ the atomic mass and the coupling strength is 
 \begin{equation}\label{eq:g}
 g=\frac{4\,\pi\hbar^2\,a}{m},
 \end{equation}
and $a$ is the scattering length. So, finally:
 \begin{equation}\label{eq:a}
 c_{s0}=\frac{\hbar}{m}\sqrt{4\pi\rho\,a}.
 \end{equation}
 It is well-known \cite{feshbachlucia} that $a$ can be modulated in time by using the dependence of the scattering length on an external magnetic field around a Feshbach resonance. The aim  is thus to achieve:
 \begin{equation}\label{eq:at}
 a(t)=a(0)(1-h(t)),
 \end{equation}
because then
\begin{equation}\label{eq:cst2}
c_s(t)=c_{s0}(1-\frac{h(t)}{2}),
\end{equation} 
(up to the first order) where 
\begin{equation}\label{eq:cs0}
c_{s0}=\frac{\hbar}{m}\sqrt{4\pi\rho\,a(0)}.
\end{equation}
To this end, we will exploit the dependence of the scattering length with an external  magnetic field
\begin{equation}\label{eq:scattlength}
a=a_{bg}(1-\frac{\omega}{B-B_0})
\end{equation}
where $a_{bg}$ is the background scattering length, $B_0$ is the value of the magnetic field at which the Feshbach resonance takes place, and $\omega$ is the width of the Feshbach resonance. 
Considering a time-dependent magnetic field
\begin{equation}\label{eq:magneticfieldt}
B(t)=B(0)(1+\delta B(t)),
\end{equation}
and with a little algebra, we can write, to the first order in $\delta B$:
\begin{equation}\label{eq:a}
a(t)=a(0) (1+\frac{B(0)\omega\delta B(t)}{(B(0)-B_0-\omega)(B(0)-B_0)}).
\end{equation}
So, if we identify:
\begin{equation}
h_+(t)=\frac{B(0)\omega\delta B(t)}{(B(0)-B_0-\omega)(B(0)-B_0)},
\end{equation}
we are simulating a gravitational wave.
Writing 
\begin{equation}\label{eq:fieldt}
\delta B(t)= \delta B \sin{\Omega\,t},
\end{equation}
the amplitude of the simulated wave is:
\begin{equation}\label{eq:amplitude}
A_+=\frac{B(0)\omega\delta B}{(B(0)-B_0-\omega)(B(0)-B_0)}
\end{equation}
and $\Omega$ is the frequency of the simulated wave. 
Taking experimental values for $B_0$ and $\omega$, \cite{feshbachlucia} and assuming that we can control the magnetic field in a $0.1 G$ scale (so we can take $\delta B\simeq B(0)\simeq 0.1 G$), we estimate the amplitude of the simulated wave as 
\begin{equation}\label{eq:amplitudesimulatedgw}
A_+\simeq 10^{-7}. 
\end{equation}
This is much larger that the gravitational waves that we expect to see in the Earth $A_+\simeq10^{-20}$, due to the fact that the Earth is far from typical sources of gravitational waves. Therefore it is interesting to think of the physical meaning of the gravitational waves that can be simulated with our techniques. 

For instance, the amplitude and frequency of the $+$ polarisation of  the gravitational wave generated by a Sun-Earth- like system are: 
\begin{equation}\label{eq:gravsunearth}
A_+=\frac{4\,G^2\,m\,M}{c^4\,R\,r}\,; \Omega=2\sqrt{\frac{G(m+M)}{r^3}}
\end{equation}
respectively, where $G$ is Newton's gravitational constant, $m$ and $M$ are the masses of the Earth and the Sun respectively, $r$ the distance between the two bodies and $R$ the distance between the detector and the centre of the mass of the system, which is assumed to be much larger than the wavelength $\lambda$ of the gravitational wave $R>>\lambda$.
Thus, if we consider the real masses of the Sun and the Earth, a distance between them of $r=10\,m$ and $R=10^7\,m$, we find $\Omega$ in the $KHz$ range -which is very convenient to generate phonons in the BEC- and the desired value of $A_+$. In \cite{ourgravwave} it is predicted that the changes in the covariance matrix of the Bogoliubov modes induced by gravitational waves of typical amplitudes $A_+<10^{-20}$ are in principle detectable. Therefore, the changes generated by a simulated wave of much larger amplitude, should be observed in an experiment with current cold-atoms technology.

\section*{Conclusions}
We have shown how to generate an artificial gravitational wave spacetime for the quantum excitations of a BEC. In the case in which there are no initial background flows, we show that the simulated ripple is obtained through a modulation of the speed of sound in the BEC. In the laboratory, this can be achieved with current technology by exploiting the dependence of the scattering length on the external magnetic field around a Feshbach resonance. With realistic experimental parameters, we find that simulated gravitational waves that can resonate with the Bogoliubov modes of the BEC. The amplitude of these artificial ripples is much larger than the typical amplitude expected for gravitational waves reaching the Earth, due to the fact that the Earth is very far from typical sources. Thus our simulated waves would mimic the waves generated by sources much closer to the BEC. This feature would enhance the effects of the ripple in the system, facilitating their detection.
The experimental test of our predictions would be a proof-of-concept of the generation of particles by   gravitational waves and would pave the way for the actual observation of \textit{real} spacetime ripples in a BEC. More generally, our low-cost Earth-based tabletop experiment will inform the whole programme of gravitational wave astronomy.

%
%


\begin{backmatter}

%
\section*{Acknowledgements}
  TB acknowledges funding from CONACYT. IF and CS acknowledge funding from EPSRC (CAF Grant No. EP/G00496X/2 to I. F.)



\newcommand{\BMCxmlcomment}[1]{}

\BMCxmlcomment{

<refgrp>

<bibl id="B1">
  <title><p>Quantum simulation</p></title>
  <aug>
    <au><snm>Georgescu</snm><fnm>I.M.</fnm></au>
    <au><snm>Ashhab</snm><fnm>S.</fnm></au>
    <au><snm>Nori</snm><fnm>F</fnm></au>
  </aug>
  <source>Rev. Mod. Phys.</source>
  <pubdate>2014</pubdate>
  <volume>86</volume>
  <fpage>153</fpage>
  <lpage>-185</lpage>
</bibl>

<bibl id="B2">
  <title><p>{Quantum simulation of the Dirac equation}</p></title>
  <aug>
    <au><snm>{Gerritsma}</snm><fnm>R.</fnm></au>
    <au><snm>{Kirchmair}</snm><fnm>G.</fnm></au>
    <au><snm>{Z{\"a}hringer}</snm><fnm>F.</fnm></au>
    <au><snm>{Solano}</snm><fnm>E.</fnm></au>
    <au><snm>{Blatt}</snm><fnm>R.</fnm></au>
    <au><snm>{Roos}</snm><fnm>C. F.</fnm></au>
  </aug>
  <source>Nature</source>
  <pubdate>2010</pubdate>
  <volume>463</volume>
  <fpage>68</fpage>
  <lpage>71</lpage>
</bibl>

<bibl id="B3">
  <title><p>Quantum Simulation of the Majorana Equation and Unphysical
  Operations</p></title>
  <aug>
    <au><snm>Casanova</snm><fnm>J.</fnm></au>
    <au><snm>Sab{\'\i}n</snm><fnm>C.</fnm></au>
    <au><snm>Le{\'o}n</snm><fnm>J.</fnm></au>
    <au><snm>Egusquiza</snm><fnm>I. L.</fnm></au>
    <au><snm>Gerritsma</snm><fnm>R.</fnm></au>
    <au><snm>Roos</snm><fnm>C. F.</fnm></au>
    <au><snm>Garc{\'\i}a Ripoll</snm><fnm>J. J.</fnm></au>
    <au><snm>Solano</snm><fnm>E.</fnm></au>
  </aug>
  <source>Phys. Rev. X</source>
  <pubdate>2011</pubdate>
  <volume>1</volume>
  <fpage>021018</fpage>
</bibl>

<bibl id="B4">
  <title><p>Encoding relativistic potential dynamics into free
  evolution</p></title>
  <aug>
    <au><snm>Sab{\'\i}n</snm><fnm>C.</fnm></au>
    <au><snm>Casanova</snm><fnm>J.</fnm></au>
    <au><snm>Garc{\'\i}a Ripoll</snm><fnm>J. J.</fnm></au>
    <au><snm>Lamata</snm><fnm>L.</fnm></au>
    <au><snm>Solano</snm><fnm>E.</fnm></au>
    <au><snm>Le\'on</snm><fnm>J.</fnm></au>
  </aug>
  <source>Phys. Rev. A</source>
  <pubdate>2012</pubdate>
  <volume>85</volume>
  <fpage>052301</fpage>
</bibl>

<bibl id="B5">
  <title><p>Quantum Simulation of Noncausal Kinematic
  Transformations</p></title>
  <aug>
    <au><snm>Alvarez Rodriguez</snm><fnm>U.</fnm></au>
    <au><snm>Casanova</snm><fnm>J.</fnm></au>
    <au><snm>Lamata</snm><fnm>L.</fnm></au>
    <au><snm>Solano</snm><fnm>E.</fnm></au>
  </aug>
  <source>Phys. Rev. Lett.</source>
  <pubdate>2013</pubdate>
  <volume>111</volume>
  <fpage>090503</fpage>
</bibl>

<bibl id="B6">
  <title><p>Relativity: special, general and cosmological, 2nd
  edition</p></title>
  <aug>
    <au><snm>Rindler</snm><fnm>W</fnm></au>
  </aug>
  <publisher>Oxford University Press</publisher>
  <pubdate>2006</pubdate>
</bibl>

<bibl id="B7">
  <title><p>The basics of gravitational wave theory</p></title>
  <aug>
    <au><snm>Flanagan</snm><fnm>EE</fnm></au>
    <au><snm>Hughes</snm><fnm>SA</fnm></au>
  </aug>
  <source>New Journal of Physics</source>
  <pubdate>2005</pubdate>
  <volume>7</volume>
  <issue>1</issue>
  <fpage>204</fpage>
</bibl>

<bibl id="B8">
  <title><p>Gravitational wave detectors</p></title>
  <aug>
    <au><snm>Aufmuth</snm><fnm>P</fnm></au>
    <au><snm>Danzmann</snm><fnm>K</fnm></au>
  </aug>
  <source>New Journal of Physics</source>
  <pubdate>2005</pubdate>
  <volume>7</volume>
  <fpage>202</fpage>
</bibl>

<bibl id="B9">
  <title><p>{Phonon creation by gravitational waves}</p></title>
  <aug>
    <au><snm>{Sab{\'{\i}}n}</snm><fnm>C.</fnm></au>
    <au><snm>{Bruschi}</snm><fnm>D. E.</fnm></au>
    <au><snm>{Ahmadi}</snm><fnm>M.</fnm></au>
    <au><snm>{Fuentes}</snm><fnm>I.</fnm></au>
  </aug>
  <source>ArXiv e-prints</source>
  <pubdate>2014</pubdate>
  <fpage>1402.7009.AcceptedinNewJournalofPhysics</fpage>
</bibl>

<bibl id="B10">
  <title><p>Quantum theory of the electromagnetic field in a variable length
  one dimensional cavity</p></title>
  <aug>
    <au><snm>Moore</snm><fnm>G. T.</fnm></au>
  </aug>
  <source>J. Math. Phys.</source>
  <pubdate>1970</pubdate>
  <volume>11</volume>
  <fpage>269</fpage>
</bibl>

<bibl id="B11">
  <title><p>Observation of the dynamical Casimir effect in a superconducting
  circuit</p></title>
  <aug>
    <au><snm>{Wilson}</snm><fnm>C. M.</fnm></au>
    <au><snm>{Johansson}</snm><fnm>G.</fnm></au>
    <au><snm>{Pourkabirian}</snm><fnm>A.</fnm></au>
    <au><snm>{Simoen}</snm><fnm>M.</fnm></au>
    <au><snm>{Johansson}</snm><fnm>J. R.</fnm></au>
    <au><snm>{Duty}</snm><fnm>T.</fnm></au>
    <au><snm>{Nori}</snm><fnm>F.</fnm></au>
    <au><snm>{Delsing}</snm><fnm>P.</fnm></au>
  </aug>
  <source>Nature</source>
  <pubdate>2011</pubdate>
  <volume>479</volume>
  <fpage>376</fpage>
  <lpage>379</lpage>
</bibl>

<bibl id="B12">
  <title><p>{Acoustic geometry for general relativistic barotropic irrotational
  fluid flow}</p></title>
  <aug>
    <au><snm>Visser</snm><fnm>M</fnm></au>
    <au><snm>Molina Paris</snm><fnm>C</fnm></au>
  </aug>
  <source>New J.Phys.</source>
  <pubdate>2010</pubdate>
  <volume>12</volume>
  <fpage>095014</fpage>
</bibl>

<bibl id="B13">
  <title><p>{Relativistic Bose-Einstein Condensates: a New System for Analogue
  Models of Gravity}</p></title>
  <aug>
    <au><snm>Fagnocchi</snm><fnm>S</fnm></au>
    <au><snm>Finazzi</snm><fnm>S</fnm></au>
    <au><snm>Liberati</snm><fnm>S</fnm></au>
    <au><snm>Kormos</snm><fnm>M</fnm></au>
    <au><snm>Trombettoni</snm><fnm>A</fnm></au>
  </aug>
  <source>New J.Phys.</source>
  <pubdate>2010</pubdate>
  <volume>12</volume>
  <fpage>095012</fpage>
</bibl>

<bibl id="B14">
  <title><p>{Testing the effects of gravity and motion on quantum entanglement
  in space-based experiments}</p></title>
  <aug>
    <au><snm>{Bruschi}</snm><fnm>D. E.</fnm></au>
    <au><snm>{Sab{\'{\i}}n}</snm><fnm>C.</fnm></au>
    <au><snm>{White}</snm><fnm>A.</fnm></au>
    <au><snm>{Baccetti}</snm><fnm>V.</fnm></au>
    <au><snm>{Oi}</snm><fnm>D. K. L.</fnm></au>
    <au><snm>{Fuentes}</snm><fnm>I.</fnm></au>
  </aug>
  <source>ArXiv e-prints</source>
  <pubdate>2013</pubdate>
  <fpage>1306.1933</fpage>
</bibl>

<bibl id="B15">
  <title><p>Analogue Gravity</p></title>
  <aug>
    <au><snm>Barcel{\'{o}}</snm><fnm>C.</fnm></au>
    <au><snm>Liberati</snm><fnm>S.</fnm></au>
    <au><snm>Visser</snm><fnm>M.</fnm></au>
  </aug>
  <source>Living Rev.~Relativity</source>
  <pubdate>2011</pubdate>
  <volume>14</volume>
  <fpage>3</fpage>
</bibl>

<bibl id="B16">
  <title><p>Sonic analog of gravitational black holes in Bose?Einstein
  condensates</p></title>
  <aug>
    <au><snm>Garay</snm><fnm>L. J.</fnm></au>
    <au><snm>Anglin</snm><fnm>J. R.</fnm></au>
    <au><snm>Cirac</snm><fnm>J. I.</fnm></au>
    <au><snm>Zoller</snm><fnm>P.</fnm></au>
  </aug>
  <source>Phys. Rev. Lett.</source>
  <pubdate>2000</pubdate>
  <volume>85</volume>
  <fpage>4643</fpage>
</bibl>

<bibl id="B17">
  <title><p>{Planck Distribution of Phonons in a Bose-Einstein
  Condensate}</p></title>
  <aug>
    <au><snm>{Schley}</snm><fnm>R.</fnm></au>
    <au><snm>{Berkovitz}</snm><fnm>A.</fnm></au>
    <au><snm>{Rinott}</snm><fnm>S.</fnm></au>
    <au><snm>{Shammass}</snm><fnm>I.</fnm></au>
    <au><snm>{Blumkin}</snm><fnm>A.</fnm></au>
    <au><snm>{Steinhauer}</snm><fnm>J.</fnm></au>
  </aug>
  <source>Physical Review Letters</source>
  <pubdate>2013</pubdate>
  <volume>111</volume>
  <issue>5</issue>
  <fpage>055301</fpage>
</bibl>

<bibl id="B18">
  <title><p>Observation of self-amplifying Hawking radiation in an analogue
  black-hole laser</p></title>
  <aug>
    <au><snm>Steinahauer</snm><fnm>J.</fnm></au>
  </aug>
  <source>Nature Phys.</source>
  <pubdate>2014</pubdate>
  <volume>10</volume>
  <fpage>864</fpage>
</bibl>

<bibl id="B19">
  <title><p>Bose Einstein Condensation in dilute gases</p></title>
  <aug>
    <au><snm>Pethick</snm><fnm>C. J.</fnm></au>
    <au><snm>Smith</snm><fnm>H.</fnm></au>
  </aug>
  <publisher>Cambridge University Press</publisher>
  <pubdate>2004</pubdate>
</bibl>

<bibl id="B20">
  <title><p>{Fermionic transport and out-of-equilibrium dynamics in a
  homogeneous Hubbard model with ultracold atoms}</p></title>
  <aug>
    <au><snm>{Schneider}</snm><fnm>U.</fnm></au>
    <au><snm>{Hackerm{\"u}ller}</snm><fnm>L.</fnm></au>
    <au><snm>{Ronzheimer}</snm><fnm>J. P.</fnm></au>
    <au><snm>{Will}</snm><fnm>S.</fnm></au>
    <au><snm>{Braun}</snm><fnm>S.</fnm></au>
    <au><snm>{Best}</snm><fnm>T.</fnm></au>
    <au><snm>{Bloch}</snm><fnm>I.</fnm></au>
    <au><snm>{Demler}</snm><fnm>E.</fnm></au>
    <au><snm>{Mandt}</snm><fnm>S.</fnm></au>
    <au><snm>{Rasch}</snm><fnm>D.</fnm></au>
    <au><snm>{Rosch}</snm><fnm>A.</fnm></au>
  </aug>
  <source>Nature Physics</source>
  <pubdate>2012</pubdate>
  <volume>8</volume>
  <fpage>213</fpage>
  <lpage>218</lpage>
</bibl>

</refgrp>
} 
\end{backmatter}


\begin{thebibliography}{20}
\ifx \bisbn   \undefined \def \bisbn  #1{ISBN #1}\fi
\ifx \binits  \undefined \def \binits#1{#1}\fi
\ifx \bauthor  \undefined \def \bauthor#1{#1}\fi
\ifx \batitle  \undefined \def \batitle#1{#1}\fi
\ifx \bjtitle  \undefined \def \bjtitle#1{#1}\fi
\ifx \bvolume  \undefined \def \bvolume#1{\textbf{#1}}\fi
\ifx \byear  \undefined \def \byear#1{#1}\fi
\ifx \bissue  \undefined \def \bissue#1{#1}\fi
\ifx \bfpage  \undefined \def \bfpage#1{#1}\fi
\ifx \blpage  \undefined \def \blpage #1{#1}\fi
\ifx \burl  \undefined \def \burl#1{\textsf{#1}}\fi
\ifx \doiurl  \undefined \def \doiurl#1{\textsf{#1}}\fi
\ifx \betal  \undefined \def \betal{\textit{et al.}}\fi
\ifx \binstitute  \undefined \def \binstitute#1{#1}\fi
\ifx \binstitutionaled  \undefined \def \binstitutionaled#1{#1}\fi
\ifx \bctitle  \undefined \def \bctitle#1{#1}\fi
\ifx \beditor  \undefined \def \beditor#1{#1}\fi
\ifx \bpublisher  \undefined \def \bpublisher#1{#1}\fi
\ifx \bbtitle  \undefined \def \bbtitle#1{#1}\fi
\ifx \bedition  \undefined \def \bedition#1{#1}\fi
\ifx \bseriesno  \undefined \def \bseriesno#1{#1}\fi
\ifx \blocation  \undefined \def \blocation#1{#1}\fi
\ifx \bsertitle  \undefined \def \bsertitle#1{#1}\fi
\ifx \bsnm \undefined \def \bsnm#1{#1}\fi
\ifx \bsuffix \undefined \def \bsuffix#1{#1}\fi
\ifx \bparticle \undefined \def \bparticle#1{#1}\fi
\ifx \barticle \undefined \def \barticle#1{#1}\fi
\ifx \bconfdate \undefined \def \bconfdate #1{#1}\fi
\ifx \botherref \undefined \def \botherref #1{#1}\fi
\ifx \url \undefined \def \url#1{\textsf{#1}}\fi
\ifx \bchapter \undefined \def \bchapter#1{#1}\fi
\ifx \bbook \undefined \def \bbook#1{#1}\fi
\ifx \bcomment \undefined \def \bcomment#1{#1}\fi
\ifx \oauthor \undefined \def \oauthor#1{#1}\fi
\ifx \citeauthoryear \undefined \def \citeauthoryear#1{#1}\fi
\ifx \endbibitem  \undefined \def \endbibitem {}\fi
\ifx \bconflocation  \undefined \def \bconflocation#1{#1}\fi
\ifx \arxivurl  \undefined \def \arxivurl#1{\textsf{#1}}\fi
\csname PreBibitemsHook\endcsname

\bibitem{review}
\begin{barticle}
\bauthor{\bsnm{Georgescu}, \binits{I.M.}},
\bauthor{\bsnm{Ashhab}, \binits{S.}},
\bauthor{\bsnm{Nori}, \binits{F.}}:
\batitle{Quantum simulation}.
\bjtitle{Rev. Mod. Phys.}
\bvolume{86},
\bfpage{153}--\blpage{185}
(\byear{2014})
\end{barticle}
\endbibitem

\bibitem{naturekike}
\begin{barticle}
\bauthor{\bsnm{{Gerritsma}}, \binits{R.}},
\bauthor{\bsnm{{Kirchmair}}, \binits{G.}},
\bauthor{\bsnm{{Z{\"a}hringer}}, \binits{F.}},
\bauthor{\bsnm{{Solano}}, \binits{E.}},
\bauthor{\bsnm{{Blatt}}, \binits{R.}},
\bauthor{\bsnm{{Roos}}, \binits{C.F.}}:
\batitle{{Quantum simulation of the Dirac equation}}.
\bjtitle{Nature}
\bvolume{463},
\bfpage{68}--\blpage{71}
(\byear{2010})
\end{barticle}
\endbibitem

\bibitem{majorana1}
\begin{barticle}
\bauthor{\bsnm{Casanova}, \binits{J.}},
\bauthor{\bsnm{Sab{\'\i}n}, \binits{C.}},
\bauthor{\bsnm{Le{\'o}n}, \binits{J.}},
\bauthor{\bsnm{Egusquiza}, \binits{I.L.}},
\bauthor{\bsnm{Gerritsma}, \binits{R.}},
\bauthor{\bsnm{Roos}, \binits{C.F.}},
\bauthor{\bsnm{Garc{\'\i}a-Ripoll}, \binits{J.J.}},
\bauthor{\bsnm{Solano}, \binits{E.}}:
\batitle{Quantum simulation of the majorana equation and unphysical
  operations}.
\bjtitle{Phys. Rev. X}
\bvolume{1},
\bfpage{021018}
(\byear{2011})
\end{barticle}
\endbibitem

\bibitem{majorana2}
\begin{barticle}
\bauthor{\bsnm{Sab{\'\i}n}, \binits{C.}},
\bauthor{\bsnm{Casanova}, \binits{J.}},
\bauthor{\bsnm{Garc{\'\i}a-Ripoll}, \binits{J.J.}},
\bauthor{\bsnm{Lamata}, \binits{L.}},
\bauthor{\bsnm{Solano}, \binits{E.}},
\bauthor{\bsnm{Le\'on}, \binits{J.}}:
\batitle{Encoding relativistic potential dynamics into free evolution}.
\bjtitle{Phys. Rev. A}
\bvolume{85},
\bfpage{052301}
(\byear{2012})
\end{barticle}
\endbibitem

\bibitem{noncausal}
\begin{barticle}
\bauthor{\bsnm{Alvarez-Rodriguez}, \binits{U.}},
\bauthor{\bsnm{Casanova}, \binits{J.}},
\bauthor{\bsnm{Lamata}, \binits{L.}},
\bauthor{\bsnm{Solano}, \binits{E.}}:
\batitle{Quantum simulation of noncausal kinematic transformations}.
\bjtitle{Phys. Rev. Lett.}
\bvolume{111},
\bfpage{090503}
(\byear{2013})
\end{barticle}
\endbibitem

\bibitem{rindlerrelativity}
\begin{bbook}
\bauthor{\bsnm{Rindler}, \binits{W.}}:
\bbtitle{Relativity: Special, General and Cosmological, 2nd Edition}.
\bpublisher{Oxford University Press}, \blocation{???}
(\byear{2006})
\end{bbook}
\endbibitem

\bibitem{reviewgravwaves}
\begin{barticle}
\bauthor{\bsnm{Flanagan}, \binits{E.E.}},
\bauthor{\bsnm{Hughes}, \binits{S.A.}}:
\batitle{The basics of gravitational wave theory}.
\bjtitle{New Journal of Physics}
\bvolume{7}(\bissue{1}),
\bfpage{204}
(\byear{2005})
\end{barticle}
\endbibitem

\bibitem{gravwavesdetectors}
\begin{barticle}
\bauthor{\bsnm{Aufmuth}, \binits{P.}},
\bauthor{\bsnm{Danzmann}, \binits{K.}}:
\batitle{Gravitational wave detectors}.
\bjtitle{New Journal of Physics}
\bvolume{7},
\bfpage{202}
(\byear{2005})
\end{barticle}
\endbibitem

\bibitem{ourgravwave}
\begin{botherref}
\oauthor{\bsnm{{Sab{\'{\i}}n}}, \binits{C.}},
\oauthor{\bsnm{{Bruschi}}, \binits{D.E.}},
\oauthor{\bsnm{{Ahmadi}}, \binits{M.}},
\oauthor{\bsnm{{Fuentes}}, \binits{I.}}:
{Phonon creation by gravitational waves}.
ArXiv e-prints,
1402--7009
(2014)
\end{botherref}
\endbibitem

\bibitem{moore}
\begin{barticle}
\bauthor{\bsnm{Moore}, \binits{G.T.}}:
\batitle{Quantum theory of the electromagnetic field in a variable length one
  dimensional cavity}.
\bjtitle{J. Math. Phys.}
\bvolume{11},
\bfpage{269}
(\byear{1970})
\end{barticle}
\endbibitem

\bibitem{casimirwilson}
\begin{barticle}
\bauthor{\bsnm{{Wilson}}, \binits{C.M.}},
\bauthor{\bsnm{{Johansson}}, \binits{G.}},
\bauthor{\bsnm{{Pourkabirian}}, \binits{A.}},
\bauthor{\bsnm{{Simoen}}, \binits{M.}},
\bauthor{\bsnm{{Johansson}}, \binits{J.R.}},
\bauthor{\bsnm{{Duty}}, \binits{T.}},
\bauthor{\bsnm{{Nori}}, \binits{F.}},
\bauthor{\bsnm{{Delsing}}, \binits{P.}}:
\batitle{Observation of the dynamical casimir effect in a superconducting
  circuit}.
\bjtitle{Nature}
\bvolume{479},
\bfpage{376}--\blpage{379}
(\byear{2011})
\end{barticle}
\endbibitem

\bibitem{matt}
\begin{barticle}
\bauthor{\bsnm{Visser}, \binits{M.}},
\bauthor{\bsnm{Molina-Paris}, \binits{C.}}:
\batitle{{Acoustic geometry for general relativistic barotropic irrotational
  fluid flow}}.
\bjtitle{New J.Phys.}
\bvolume{12},
\bfpage{095014}
(\byear{2010})
\end{barticle}
\endbibitem

\bibitem{liberati}
\begin{barticle}
\bauthor{\bsnm{Fagnocchi}, \binits{S.}},
\bauthor{\bsnm{Finazzi}, \binits{S.}},
\bauthor{\bsnm{Liberati}, \binits{S.}},
\bauthor{\bsnm{Kormos}, \binits{M.}},
\bauthor{\bsnm{Trombettoni}, \binits{A.}}:
\batitle{{Relativistic Bose-Einstein Condensates: a New System for Analogue
  Models of Gravity}}.
\bjtitle{New J.Phys.}
\bvolume{12},
\bfpage{095012}
(\byear{2010})
\end{barticle}
\endbibitem

\bibitem{salelites}
\begin{botherref}
\oauthor{\bsnm{{Bruschi}}, \binits{D.E.}},
\oauthor{\bsnm{{Sab{\'{\i}}n}}, \binits{C.}},
\oauthor{\bsnm{{White}}, \binits{A.}},
\oauthor{\bsnm{{Baccetti}}, \binits{V.}},
\oauthor{\bsnm{{Oi}}, \binits{D.K.L.}},
\oauthor{\bsnm{{Fuentes}}, \binits{I.}}:
{Testing the effects of gravity and motion on quantum entanglement in
  space-based experiments}.
ArXiv e-prints,
1306--1933
(2013)
\end{botherref}
\endbibitem

\bibitem{analoguereview2011}
\begin{barticle}
\bauthor{\bsnm{Barcel{\'{o}}}, \binits{C.}},
\bauthor{\bsnm{Liberati}, \binits{S.}},
\bauthor{\bsnm{Visser}, \binits{M.}}:
\batitle{Analogue gravity}.
\bjtitle{Living Rev.~Relativity}
\bvolume{14},
\bfpage{3}
(\byear{2011})
\end{barticle}
\endbibitem

\bibitem{luisblackhole}
\begin{barticle}
\bauthor{\bsnm{Garay}, \binits{L.J.}},
\bauthor{\bsnm{Anglin}, \binits{J.R.}},
\bauthor{\bsnm{Cirac}, \binits{J.I.}},
\bauthor{\bsnm{Zoller}, \binits{P.}}:
\batitle{Sonic analog of gravitational black holes in bose?einstein
  condensates}.
\bjtitle{Phys. Rev. Lett.}
\bvolume{85},
\bfpage{4643}
(\byear{2000})
\end{barticle}
\endbibitem

\bibitem{megamind}
\begin{barticle}
\bauthor{\bsnm{{Schley}}, \binits{R.}},
\bauthor{\bsnm{{Berkovitz}}, \binits{A.}},
\bauthor{\bsnm{{Rinott}}, \binits{S.}},
\bauthor{\bsnm{{Shammass}}, \binits{I.}},
\bauthor{\bsnm{{Blumkin}}, \binits{A.}},
\bauthor{\bsnm{{Steinhauer}}, \binits{J.}}:
\batitle{{Planck Distribution of Phonons in a Bose-Einstein Condensate}}.
\bjtitle{Physical Review Letters}
\bvolume{111}(\bissue{5}),
\bfpage{055301}
(\byear{2013})
\end{barticle}
\endbibitem

\bibitem{jeffblackhole}
\begin{barticle}
\bauthor{\bsnm{Steinahauer}, \binits{J.}}:
\batitle{Observation of self-amplifying hawking radiation in an analogue
  black-hole laser}.
\bjtitle{Nature Phys.}
\bvolume{10},
\bfpage{864}
(\byear{2014})
\end{barticle}
\endbibitem

\bibitem{pethicksmith}
\begin{bbook}
\bauthor{\bsnm{Pethick}, \binits{C.J.}},
\bauthor{\bsnm{Smith}, \binits{H.}}:
\bbtitle{Bose Einstein Condensation in Dilute Gases}.
\bpublisher{Cambridge University Press}, \blocation{???}
(\byear{2004})
\end{bbook}
\endbibitem

\bibitem{feshbachlucia}
\begin{barticle}
\bauthor{\bsnm{{Schneider}}, \binits{U.}},
\bauthor{\bsnm{{Hackerm{\"u}ller}}, \binits{L.}},
\bauthor{\bsnm{{Ronzheimer}}, \binits{J.P.}},
\bauthor{\bsnm{{Will}}, \binits{S.}},
\bauthor{\bsnm{{Braun}}, \binits{S.}},
\bauthor{\bsnm{{Best}}, \binits{T.}},
\bauthor{\bsnm{{Bloch}}, \binits{I.}},
\bauthor{\bsnm{{Demler}}, \binits{E.}},
\bauthor{\bsnm{{Mandt}}, \binits{S.}},
\bauthor{\bsnm{{Rasch}}, \binits{D.}},
\bauthor{\bsnm{{Rosch}}, \binits{A.}}:
\batitle{{Fermionic transport and out-of-equilibrium dynamics in a homogeneous
  Hubbard model with ultracold atoms}}.
\bjtitle{Nature Physics}
\bvolume{8},
\bfpage{213}--\blpage{218}
(\byear{2012})
\end{barticle}
\endbibitem

\end{thebibliography}
 \end{document}